\documentclass{article}
\usepackage{tabularx}
\usepackage{threeparttable}
\usepackage{amsthm}
\usepackage{ijcai19}

\usepackage{times}
\usepackage{soul}
\usepackage{url}
\usepackage[hidelinks]{hyperref}
\usepackage[utf8]{inputenc}
\usepackage[small]{caption}
\usepackage{graphicx}
\usepackage{rotating}
\usepackage{amsmath}
\usepackage{booktabs}
\usepackage{algorithm}
\usepackage[noend]{algpseudocode}
\title{Towards Algorithmic Transparency: A Diversity Perspective}

\author{
Fausto Giunchiglia$^1$\footnote{Contact Author}\and
Jahna Otterbacher$^2$\and
Styliani Kleanthous$^2$ \and \\
Khuyagbaatar Batsuren$^{1}$\and
Veronika Bogina$^3$ \and \\
Tsvi Kuflik$^3$ \And
Avital Shulner Tal$^3$\\
\affiliations
$^1$Department of Information Engineering and Computer Science, University of Trento, IT \\
$^2$School of Pure and Applied Sciences, Open University of Cyprus, CY\\
$^3$Department of Information Systems, The University of Haifa, IL\\
\emails
\{fausto.giunchiglia, k.batsuren\}@unitn.it,
\{jahna.otterbacher, styliani.kleanthous\}@ouc.ac.cy,\\ 
\{sveron, avitalshulner\}@gmail.com,
tsvikak@is.haifa.ac.il
}

\begin{document}

\maketitle

\begin{abstract}
As the role of algorithmic systems and processes increases in society, so does the risk of bias, which can result in discrimination against individuals and social groups. Research on algorithmic bias has exploded in recent years, highlighting both the problems of bias, and the potential solutions, in terms of \textit{algorithmic transparency} (AT). Transparency is important for facilitating fairness management as well as explainability in algorithms; however, the concept of \textit{diversity}, and its relationship to bias and transparency, has been largely left out of the discussion. We reflect on the relationship between diversity and bias, arguing that diversity drives the need for transparency. Using a perspective-taking lens, which takes diversity as a given, we propose a conceptual framework to characterize the problem and solution spaces of AT, to aid its application in algorithmic systems. Example cases from three research domains are described using our framework. 
\end{abstract}

\section{Introduction}

Algorithmic systems are undeniably socio-technical in nature \cite{barocas-hardt-narayanan}. In machine learned systems, training and evaluation datasets aim to capture aspects of the state-of-the-world, and learning mechanisms are applied to create models based on them. Thus, systems naturally reflect the biases of the societies in which they are developed. Some lead to discrimination, as in the much discussed case of the COMPAS system for predicting recidivism \cite{angwin2016machine}, or computer vision applications that infer a depicted person's demographic characteristics \cite{buolamwini2018gender}. However, the aim of developing algorithmic systems is to introduce them into social contexts, with the intent that they will serve their intended purposes, without harming people. 

Unfortunately, we often ignore the intentional or unintentional consequences that the system will have when introduced into a social context \cite{selbst2019fairness}. When designing a system, we have an implicit model of its users, based on the development process followed. Yet, when we deploy the system, we cannot anticipate who will use it and how. The actual users of our system will be diverse, and likely different, than the audience we initially had in mind. In other words, this exposure to \textit{diversity} -- in a world globalized by the Internet -- has had a profound effect on the development of information systems.

A decade ago, the issue of diversity was much less critical, as the space and time limitations shielded us from it. However, this is no longer the case. All of us -- and the systems we develop --  are exposed to new diversity on a daily basis, manifested in language, data and knowledge. 
Arguably, the diversity of users of networked, intelligent algorithmic systems has exacerbated the issue of social and cultural bias. A system may show behaviors that deviate from what users expect, or what they consider to be normal with respect to their own context and perspective. When deviations in system behavior are perceived, this then leads to discussions of whether the system is behaving in a manner that is fair. However, there is no single standard to which we can compare the behaviors of a given system; with a globalized user base, what is ``normal" depends on many contextual factors, including one's socio-cultural environment and the prevailing values in a society \cite{dignum2017responsible}.

Transparency has long been recognized as a desirable property of a system \cite{tyugu2006understanding}. Recently, researchers have emphasized its importance in ensuring that systems can be held accountable for harmful behaviors or the perception of unfairness \cite{diakopoulos2016accountability,datta2016algorithmic}. Yet research on algorithmic biases and algorithmic transparency (AT) is dispersed across communities, with little consensus on definitions/approaches (see for example, the ``21 definitions of fairness" \cite{narayanan2018translation} and ``50 Years of Test (Un)fairness: Lessons for Machine Learning" \cite{Hutchinson:2019:YTF:3287560.3287600}).

Danks and London \cite{danks2017algorithmic}, citing the need for a conceptualization of algorithmic system bias, put forth a taxonomy, including biases of a computational origin, as well as those arising from inappropriate use of a system. They detailed five sources: i) training data, ii) algorithmic focus (i.e., differential usage of attributes in the training data), iii) processing (e.g., use of a statistically biased estimator in a model), iv) transfer context (i.e., application in a context differing from the one for which the system was developed) and v) interpretation bias (i.e., user misinterpretation of the system output).

We argue that the notion of \textit{diversity} is key to understanding the above sources of bias. Since diversity is inevitable, perspective taking \cite{galinsky2000perspective} (i.e., interpreting the world in someone else's shoes), can be a tool for determining when bias is problematic, and for whom. Thus, we motivate a ``diversity perspective," taking into account the perspective taking work from psychology \cite{davis1983measuring}, bringing it into the discussion on algorithmic system bias and transparency. Our intent is not to advocate for a ``view from nowhere" but rather, to highlight the fundamental relationship between diversity, bias and transparency. In a globalized world, diversity drives the need for AT; without transparency one cannot detect or understand system biases. The current paper provides a different view, and continues the discussion initiated by Hutchinson and Mitchell \cite{Hutchinson:2019:YTF:3287560.3287600}, on Fairness in ML systems, with the aim of minimizing the gap between theory on AT and its application in intelligent algorithmic systems.

\section{Diversity and Bias}

Diversity is considered by many disciplines to be a positive attribute. Nelson \cite{Nelson:2014:DD:2684442.2597886} stresses that diverse teams -- in terms of skill set, education, work experiences, perspectives on a problem, cultural orientation, gender, etc. -- produce better results compared to homogeneous teams. Similarly, in \cite{galinsky2015maximizing}, the authors emphasize that the USA's success as a nation is strongly based on the diversity of its immigrants. However, Nelson also stresses that stereotypes (e.g., based on race or gender) that we formulate in societies are potential threats to a diverse team's success. This is due to expecting certain outputs and/or behaviors from particular social groups (e.g., men vs. women, African Americans vs. Caucasians) and when a deviation occurs, unfair treatment can cause conflicts within the team. 

Galinsky et al. advocate that transparency in work practices can help a diverse society to thrive \cite{galinsky2015maximizing} and minimize conflicts due to biases. As mentioned in Section 1, the Internet today, (along with its applications, such as social networks, search engines, recommender systems, decision support systems, etc.), represents a diverse community that we cannot ignore. As scientists and researchers of Web, social, and/or intelligent algorithmic systems, we face diversity in many forms: the diversity in data on the Web, the diversity of humans involved in developing and using a system, diversity in the output/information delivered to the user, to name just a few factors. 

There is growing acknowledgment that promoting \textit{transparency} is a direction towards minimizing the unwanted side-effects (such as stereotyping and biases) of diversity in data, humans and output, consequently promoting fairness. Baeza-Yates \cite{baeza2018bias}, in his article concerning the different types of biases on the Web, concludes by stating that we can only reduce bias if we are aware that it exists. Thus, developers\footnote{We use the term ``developers" loosely here, to mean anyone involved in the process, not only ML practitioners.} of algorithmic systems need to be conscious of the diversity -- based on culture, race, gender, knowledge, etc. -- of the potential consumers of their system's output, making their systems transparent at different levels.\footnote{This will be explored in detail in Section 4.} In other words, they need to first recognize that there are alternative perspectives, in addition to their own.

\subsection{Perspective taking and diversity}

Researchers in psychology have emphasized the importance of perspective taking in occasions where significant \textit{diversity} exists, (e.g., differences in the knowledge, political views, or cultural backgrounds of participants), aiming to reduce prejudice \cite{shih2009perspective}, minimize bias and stereotyping \cite{todd2011perspective} and increase in- and out-group empathy \cite{galinsky2000perspective,shih2009perspective,vescio2003perspective}. Perspective-taking studies require one person to view a situation or behavior from another's point of view \cite{galinsky2000perspective}; hence, the relevance to \textit{diversity} and bias research. 

Studies involving culturally diverse groups of people have illustrated the power of perspective taking. In their studies, Vescio and colleagues \cite{vescio2003perspective} found that individuals engaged in perspective-taking endorsed pro-African American attitudes, minimizing other individuals' bias towards African Americans in the group. Unlike humans, systems do not have the ability to change their perspective towards a situation, a statement or a belief. However, the humans involved in each step of the development of an algorithmic system (from requirement analysis to input training data and algorithm development) do have the ability, and need to consider perspective-taking as one approach towards Algorithmic Systems Transparency. 

In their work, Shih et al.\cite{shih2009perspective} examined participants' empathic feelings towards an Asian character, depicted in a scenario in which he tried to overcome a societal stereotype. Overall, the study involved Caucasian, African American, Hispanic and other non-Asian participants. It was found that participants who engaged in perspective-taking showed empathic feelings and liking for the character, thus illustrating that perspective-taking can reduce prejudice towards out-group members. Similarly, Galinsky \cite{galinsky2000perspective}, in three experimental studies, found that allowing participants to see a different perspective, decreased stereotypic biases toward out-group members, by enabling participants to associate themselves with the subject. The results showed that participants demonstrated both in-group as well as out-group favoritism. In summary, perspective taking research includes a number of studies consistent with the above findings, underscoring its promising potential application in the context of FATE (fairness, accountability, transparency and ethics) research. 


Therefore, we adopt a perspective-taking lens. We  propose a framework for discussing and understanding bias in algorithmic systems, considering variables such as the cultural, language, knowledge and life-experience \textit{diversity} of the users, the developer and the observer/researcher. 
Within data and systems, such diversity results in the co-existence of competing and/or contradictory statements, some of which may be non-factual or referring to opposing beliefs or opinions \cite{giunchiglia2012domains}. A system's user base may be global, serving individuals who perceive the world differently and do not interpret system behaviors in the same way. Thus, algorithmic systems must take \textit{diversity} into account to enhance user experience \cite{kunaver2017diversity,gu2017diversity}. 
Given the challenges, our next steps are to relate the notions of diversity and bias.

\subsection{Representing knowledge in the world}
Diversity is manifested in the ways that implicit knowledge is represented, even when we limit the discussion to the ``factual" aspects of knowledge \cite{jovchelovitch2002re}. The same entity (object/person/event) may be described in infinite ways across observers, varying by community, culture and language, or even life experience \cite{galinsky2000perspective}. When we describe an entity, we choose the properties to use, which according to our background, will best characterize that entity. In other words, the resulting description can be a person's perspective towards a situation or an entity. These properties define the space, $S$, over which we shall eventually measure bias. 

To illustrate this, consider the entity \textit{snail}. As snails have been common in European and Mediterranean cuisine for thousands of years, individuals from such cultures would likely use the property ``food" when describing snails. In contrast, a person from East Asia might relate snails to beauty products, rather than food. Still, the strength of association may vary by the observer's gender or age. This diversity of perspective is reflected in the data used to train systems. The text and multimedia shared via the Web, often used in training corpora, reflects our perspectives and experiences. Similarly, crowdsourcing often involves asking workers of various backgrounds to annotate or judge an entity relying on their own understanding of the world. Thus, embedding in the data one's perspective, which might encompass stereotypes and biases \cite{duan2020does}. 

\subsection{Defining an unbiased point of reference}
Another aspect of diversity that we need to consider, concerns the choice of the reference or standard: the unbiased point, $O$. According to the Oxford dictionary, a standard is ``something used as a measure, norm, or model in comparative evaluations." However, the choice of the reference is not common across all observers. Specifically, Jones and Nisbett noted that people can process information in different ways due to their divergent perspectives \cite{jones1987actor}. 

Continuing on the previous example, many individuals raised in the US or UK have a strong aversion to snails as food or otherwise.\footnote{As evidenced by the popular saying that boys are ``made of snails and puppydog tails," while girls are ``made of sugar and spice and everything nice."} Those who are Jewish or Muslim often share this aversion, as snails are neither kosher nor Halal. For such individuals, the unbiased area in the hyperspace representing the entity \textit{snail} will allow for little variance with respect to dimensions such as ``food" or ``pet," as compared to individuals of other backgrounds.

\subsection{Measuring bias}
Diversity also has a relation to bias, in the choice of the metric $M$, to express the deviation from a given point, to the reference. In contrast to the statistical tests discussed in \cite{Hutchinson:2019:YTF:3287560.3287600} for measuring bias, here we assume that an entity (incoming observation)
and one's reference, are represented as vectors in an $n$-dimensional space, where dimensions represent properties used to describe the entity. One could measure the distance using any number of measures (e.g., Euclidean, Cosine or Manhattan distance). However, distance measures have properties that make them more/less informative given various considerations (e.g., dimensionality). This affects perception of the deviation and thus, whether the observation is ``biased." Table~\ref{tab:notation} summarizes the formal notation.

\begin{table*}[ht]
    \centering
    \tiny
    
  \begin{tabular}{l|c}
    \hline
     \textbf{Definition} & \textbf{Notation} \\
    \hline
      Space &  $S$ \\
            \hline
     Reference point & $O$ \\ 
            \hline 
  Metric: how the distance between a given point and $O$ is quantified within $S$ & $M$ \\
            \hline 
     Individual & $i$    \\
            \hline
    Algorithmic Bias    &   AB    \\
            \hline
    The system developer &  $D$ \\
            \hline
    The system observer & $V$  \\
        \hline
     Context in which the facts of the world are represented & $C$    \\
            \hline
    Individual $i$ perceives the world with respect to her own context & $C_i$ \\
            \hline
  $S_i$ is the metric space in which $i$ interprets the world, and $O_i$ represents what is ``normal" for $i$&   $C_i =<S_i,O_i>$ \\
            \hline
A statement concerning the world & $a$ \\
            \hline
Statement $a$ belongs to a specific context $C$ & $a \in C$ \\
            \hline
The Bias Space, the  context  in  which  Bias  is  perceived  and  measured & $B=<C,M>$ \\
            \hline
Individual $i$ observes $a$, from her own context, $C_i$.  & $PB_i(a)=||a - O_i||$ \\ The bias of the statement $a$ is given by its distance from the individual's reference point $O_i$,&  \\ with respect to her metric for measuring distance in $S$. The Perceived Bias of $i$ is:&  \\
            \hline
 AB depends on the reference contexts of two parties & $D$, and $V$.\\
            \hline
By default, the reference context of the system is that of its developer & $C_D = <S_D,O_D> \neq C_V$ \\
            \hline
Perceived Algorithmic Bias & $PAB_V(a)=||a - O_V|| \geq 0$ \\
            \hline
Average System Bias &  $Mean_{PAB_V}=\frac{1}{N}\sum_{k=1}^NPAB_V(a_k)$ \\
            \hline
    \end{tabular}
    \caption{Formal representation.}
    \label{tab:notation}
\end{table*}

\subsection{Defining and measuring algorithmic bias}
We have seen that diversity relates to a general notion of bias in terms of: i) how facts of the world are represented in data and information; 
ii) the standard against which any incoming observation will be compared; and 
iii) how the deviation between the observation and the standard is measured. 

With the above in mind, we provide the following definitions. In Table~\ref{tab:notation}, the formal notation of the below definitions are provided.

\emph{Reference Context.} Before moving onto formally defining algorithmic bias, we need to acknowledge that each individual observes the world context differently or from a different perspective. Hence, context is an important part of understanding diversity and bias. As a result, the contextual reference point (individual perspective) for what is ``normal" differs from person to person. For example, take $C$ to be the context in which the facts of the world are represented. Individual $i$ perceives the world with respect to her own context, $C_i$, where $C_i =<S_i,O_i>$. $S_i$ is the metric space in which $i$ interprets/describes the world, and $O_i$ represents what is ``normal" for her. From here on Context will be used to represent a person's individual perspective.

\emph{Bias.} 
Whether characterizing a situation, an event or an object, we are making a statement about the state of the world. Bias is a property of a given statement. However, this bias is independent of whether the statement is true or false, and depends strongly on the context that intermingles the individual's point of view, with her means of making comparisons (i.e., distance metric) (see formalization in Table~\ref{tab:notation}). Assume a statement $a$ concerning the world, where $a \in C$. The Bias Space, the context in which Bias is perceived and measured, is $B=<C,M>$.  When we observe a situation, an event or an object in the world we do so from our own context. Thus, the deviation of a statement (measured distance), used for describing what we experience at a given point in $S$, from the reference point $O_i$ we have, can be considered as the bias of the statement. Therefore, the Perceived Bias of $i$ is $PB_i(a)=||a - O_i||$.

 \emph{Algorithmic Bias.} Systems, like people, also make statements describing the state of the world. Algorithmic bias (AB) is the bias generated by an algorithmic system. Extrapolating from the above definition of bias, algorithmic bias depends on the reference contexts of two parties: the developer $D$, and the system observer $V$.\footnote{The observer may or may not be a user of the system. We shall return to this point.} 
 
 By default, the reference context of the system is that of its developer. Thus, we often have that the context of the developer will be different from the context of the system observer $C_D = <S_D,O_D> \neq C_V$. This is a key cause of algorithmic bias: the system is built under the developer's context $C_D$, but its behaviors are interpreted under a different reference context $C_V$. 
 
 For example, when a developer is adding specific rules in an interactive dating system (e.g., how recommendations are made to a user), the developer is acting according to her own reference context ($C_D$). When a researcher (i.e., a system observer) is auditing the system's output given a specific input, then the researcher will interpret this based on his own context ($C_V$). 

  \emph{Measuring Algorithmic Bias.}
  We can measure the algorithmic bias of a given system by taking into account the deviation of the system-generated statement, $a$, from the observer's context or viewpoint, $C_V$, and thus, bias space, $B_V$. In this case, the Perceived Algorithmic Bias will be $PAB_V(a)=||a - O_V|| \geq 0$. It is important to note that from the developer's perspective, the algorithmic bias of the system generated statement will be zero ($PAB_D(a)=||a - O_D||=0$) assuming that she acts in good faith. Continuing on the dating system example above, if we assume that the developer did not inject any intentional bias within the system rules (acting in good faith), the perceived bias of a system output $PAB_D(a)$ will be zero. However, when an observer is interacting with the system, her perceived bias for a system output $PAB_V(a)$ will be greater than zero, given that she has a diverse background, as compared to the developer.
 

  \emph{Measuring Average System Bias.} 
Finally, it should be noted that algorithmic bias can also be measured across a representative sample of $N$ statements, $a_1, a_2,  ... a_N$, generated by the system. For instance, the mean system bias might be calculated, again from the perspective of the observer, as  $Mean_{PAB_V}=\frac{1}{N}\sum_{k=1}^NPAB_V(a_k)$. As will be shown in Section 3, researchers are often aiming to characterize the average bias of a given algorithmic system, rather than whether or not an individual algorithmic statement is biased.

\subsection{Diversity and value judgments}
We have seen that algorithmic systems make \textit{value judgments} concerning the world, which may or may not align with those of an observer of the system. An algorithm is an artefact produced by a human developer, on the basis of some reference context ($C_D$). As mentioned in the introduction, while the developer has an implicit model of the user during the development process, in reality, users will be diverse and perhaps unexpected. Any observer of the system's behavior brings in a second reference context ($C_V$), which typically differs from that of the developer, and it is under this reference context and bias space ($B_V$), where the evaluation of the system behaviors takes place. 

It must be noted that the determination of the reference context is also a value judgment, because it defines not only how we perceive the world (i.e., the space in which we characterize what we see, $S$) but also what we perceive as being expected, normal, or unbiased ($O$). One's reference context - and thus, bias space ($B$) -  is a result of her culture and upbringing, and may change (at least a bit) over time, with life experience. Thus, it is clear that the diversity of the world, and in particular, the differences between the reference contexts of people, is the root of algorithmic bias.

\section{Problem Space: Bias in Systems and Processes}
Now we examine how diversity and bias manifest and are measured within an algorithmic system. To this end, we review examples of AT research across three domains, using the definitions of Section 2. First, we provide a general characterization of algorithmic systems and their macro components, as well as of the role of the researcher as the system observer.
A basic architecture is provided in Figure~\ref{fig:AIsystem}. First, the system receives input (I) for an instance of its operation; its operational component (i.e., algorithmic model - (M)) performs some computation based on the inputs provided and produces an output (O). The model learns from a set of observations of data (D) from the problem domain. It may receive constraints from third party actors (T), and/or fairness criteria (F), which modify the operation of the algorithmic model (M). 

\subsection{Researcher as observer}
Recall that Algorithmic Bias (AB) depends on the reference contexts of the developer and the observer and this reference context will define their perspective towards the system's output. While the context of the developer $C_D$ may be unknown, the context of the observer $C_V$ is known or implied by the manner in which the study is conducted, and serves as the context for the evaluation. Thus, AT research characterizes the average algorithmic bias AB of a system, as perceived by the researcher/observer, $PAB_V(a)=||a - O_V||$.

\begin{figure}[ht]
\centering
\includegraphics[width=3.2in]{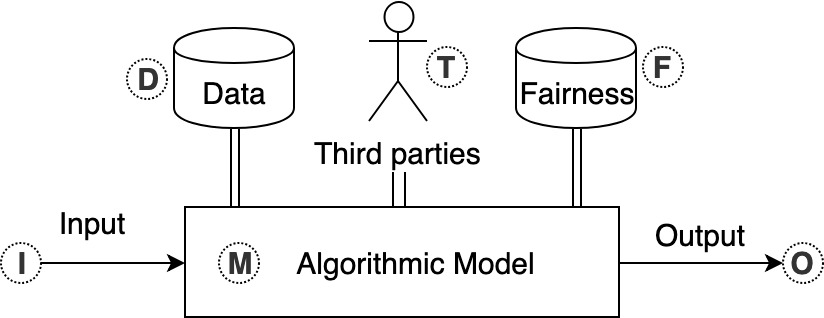}
\caption{General architecture of an algorithmic system.\label{fig:AIsystem}}
\end{figure}

With the above in mind, we analyze examples of the problem space, as presented in publications from three domains - text classification, search engines and recommender systems, leaving the discussion of the solution space for Section 4. When considering the problem space of each example we: 

\begin{itemize}
    \item Characterize the algorithmic system addressed, in relation to Figure~\ref{fig:AIsystem}. 
    \item Identify the relevant dimension(s) of diversity.
    \item Detail the manner in which $PAB_V$ is calculated.
\end{itemize}

\subsection{Text classification}
\emph{Problem.} Given an input text (I), the goal is to assign one or more appropriate classes (O) that describe some attribute of the text. In the case of a classifier based on supervised learning (M), such as those in the examples below, the necessary data for training the model is a corpus of text from the domain of interest, in which each observation/text has been labelled with the correct class(es) (D). 

\emph{Example 1 [TC1].} Dixon and colleagues \cite{dixon2018measuring} trained a binary classifier (M) to label Wikipedia comments as being toxic/not toxic (O), based on the words in the textual comment (I). The \textit{diversity dimension} of interest was minority status, including groups based on sexual orientation and religious affiliation, which were flagged by the use of sensitive identity terms (e.g., gay, black, atheist, Muslim) that appeared in the training corpus (D). 

The concern was that sensitive words were associated more frequently with examples of ``toxic" rather than ``not toxic" comments, resulting in unintended biases in the classifier. Therefore, the authors proposed balancing the toxic/not toxic examples in the dataset (F). Average perceived algorithmic bias ($PAB_V$) was calculated based on error rate metrics (i.e., comparing the rates of false positives and negatives) when classifying comments containing sensitive words versus those not containing such terms. Thus, the reference point of the observer($O_V$) was the classification error rate for texts containing no sensitive terms. 

\emph{Example 2 [TC2].} Shen and colleagues presented a sentiment analysis scenario \cite{shendarling}. Social media texts (I) were labeled as having  positive/neutral/ negative affect (O). Various black-box algorithms (M) were trained on corpora from social media (D). The \textit{diversity dimension} of interest was race. Specifically, the authors were concerned about \textit{stylistic bias}, such that texts containing linguistic markers of African-American English (AAE), were often mislabeled as negative. They proposed to ``neutralize" incoming texts, such that the algorithms would analyze them like a text of comparable context, but without sensitive terms (F).

A regression model was used to calculate average perceived algorithmic bias ($PAB_V$). Specifically, each sentiment algorithm was used to score the original datasets, as well as datasets in which sensitive words were neutralized. The regression model related the two sets of sentiment scores, such that average $PAB_V$ represented the average change in sentiment scores (i.e., change in regression coefficient). In other words, $O_V$ referred to the sentiment scores of texts in which markers of AAE were absent. 

\subsection{Search engines}
\emph{Problem.} Given an input query from a user (I), a search engine returns a ranked list of documents (O), meant to be relevant to the user's information need. The algorithmic models (M) behind modern proprietary search engines such as Google or Bing are difficult to characterize, not only because they are trade secrets, but also because of their complexity. The data used to train the models (D) likely consists of a combination of curated relevance datasets, datasets collected from ``the wild," as well as user history and profile, etc.  

\emph{Example 1-2 [SE1,SE2].} Two recent studies addressed the diversity dimension of \textit{gender}, in proprietary image search engines, considering search results (O) returned in response to queries (I) concerning the professions \cite{kay2015unequal} as well as character traits \cite{otterbacher2017competent}. Fairness constraints (F) were not suggested; rather, the aim was to document bias to raise user awareness. 

In \cite{kay2015unequal}, the reference point of the observer,  $O_V$, was derived from US labor statistics on a given profession (i.e., the gender distribution of workers in the profession). The authors noted that the choice of $O_V$ was not neutral, but rather, reflected the biases of the offline world. This was used as a benchmark of gender bias in the search engine results (presented to US-based users). Average $PAB_V$ was computed by comparing the online versus offline gender distribution in retrieved images, across a set of profession-related queries.

In contrast, to compare the gender distribution of images retrieved for a given character trait query (e.g., ``sensitive person"), no offline reference point was cited in \cite{otterbacher2017competent}. Instead, the authors compared the images retrieved on a given query, across four search engine markets (US, UK, South Africa and India). Average pairwise deviation was examined between the four markets, by comparing the gender distributions in images retrieved across a large set of queries demonstrating different observer's reference point $O_V$.

\begin{table*}[ht]
    \begin{tabular}{l|c|c|c}
    \hline
     \textbf{Diversity dimension} & \textbf{Representation ($S$)} & \textbf{Reference point ($O_V$)} & \textbf{Distance metric ($M$)} \\
    \hline
     TC1: Minority status &Word vector &Classification performance on & Error rate equality \\
                            &  &text without sensitive words &difference \\ 
            \hline 
     TC2: Race & Word vector    &Classification performance on &Change in sentiment scores \\
               &           &text without sensitive words & \\
            \hline
        SE1: Gender &Gender distribution in  &Gender distribution per &Distance between online/ \\
                   &images retrieved       &labor statistics &offline gender distributions \\
            \hline 
    SE2: Gender & Gender distribution in & None assumed&Pairwise differences \\
            & images retrieved     &           & between search markets\\
    \hline 
     SE3: Information&Distribution of URLs &``Fair results" set &One minus the similarity \\ 
                     & retrieved       &from multiple engines &between fair/empirical results \\
    \hline 
     RS1: Gender, race &Feature vector &Ranking of worker with &Coefficient on diversity attributes  \\
                   &  (text, rating, rank)             & similar history in system & in model to infer rank\\
    \hline
     RS2: Credibility &Rating, source &Rating at other platforms & Differences across platforms\\
         \hline
    \end{tabular}
    \caption{Comparison of problem spaces in AT case studies across domains.}
    \label{tab:problem-spaces}
\end{table*}

\emph{Example 3 [SE3].} Mowshowitz and Kawaguchi demonstrated a method for measuring search engine bias in a proprietary search engine \cite{mowshowitz2005measuring}. Thus, a fairness constraint (F) was not imposed on the system.  For a large set of user-generated search queries (I), they created a ``fair results set," consisting of results retrieved for a given query, across a number of alternative search engines. The diversity dimension of interest is \textit{information diversity}. Put another way, $O_V$ expects adequate diversity in search results, regardless of the topic of the query. Thus, average $PAB_V$ was calculated, based on the deviation of the search results given by the engine being evaluated (O), from the search results of the ``fair results set."

\subsection{Recommender systems}
\emph{Problem.} Given the profile of a user and an expressed need (i.e., query) (I), the system generates a ranked list of recommendations (O), deemed to be most compatible with the user's interests/needs. The system's model (M) is trained on datasets (D) capturing all users' interactions with the system. Additional constraints may be added by third parties (T) who interact with / take decisions in the system.

\emph{Example 1 [RS1].} A recent study examined gender and racial biases in two freelance marketplaces, TaskRabbit and Fiverr \cite{hannak2017bias}. In both systems, users receive ranked lists of candidates (O) in response to search terms concerning small jobs (I). Third parties leave textual comments as well as ratings for candidates (T), who have completed previous jobs. The assumption was that candidates with similar qualifications and history of experience, should receive similar rankings, regardless of their gender/race. 

The authors considered the feedback of candidates from other users, the language of the textual reviews on candidates' work (i.e., the use of positive/negative adjectives), and the candidates' positions in resulting rankings for a given job. Regression models were used to study the relationship of gender/race with these variables. The average system bias was measured in terms of the coefficients on gender/race (statistical significance, effect size). Ideas were put forward, such as re-ranking candidates to ensure fair treatment (F).

\emph{Example 2 [RS2].} Another example \cite{eslami2017careful}, presented a cross-platform audit of hotel recommendations. The systems studied return a ranked list of recommendations (O), in response to a user's profile and search terms (I), taking into consideration others' ratings/reviews (T). \textit{Credibility} was the dimension of interest, as the primary concern was that the system Booking.com skewed users' ratings of hotels, as compared to other systems. The authors compared customer ratings of over 1.500 hotels, by taking the average difference in ratings between any two platforms, and testing for statistical significance. Specific fairness constraints (F) were not mentioned; however, the authors noted that users often raise awareness of biases through textual reviews of the hotels.

\subsection{Summary of observations}
By design, all algorithmic models express intentional bias (e.g., in a search engine, a preference for ``relevant" content over that which is deemed less relevant). However, some algorithmic biases are unintended and potentially problematic, such as those examined above. Table~\ref{tab:problem-spaces} summarizes the problem spaces in these cases. As can be seen, the diversity dimensions examined reflect the potential for algorithmic biases to result in harm, such as discrimination against particular social groups (based on characteristics such as race, gender, religion or sexual orientation), or providing information to users that is not balanced or credible. 

Another observation from Table~\ref{tab:problem-spaces} is that, while the representation of knowledge ($S$) used in a given system, as well as the distance metric used ($M$), are dependent on the domain and the problem at hand, it is interesting to note a commonality in the choice of the reference point ($O_V$). In particular, all reflect an attempt to find a ``cultural consensus" on the baseline, either through the use of crowd wisdom obtained through open Web and social media (e.g., TC1, TC2, SE3), comparable data within the same or another system (RS1, RS2) or official government statistics (SE1).

In all of the above examples, we can appreciate the importance of perspective taking as a way to acknowledge the \textit{diversity}, that exists in the data, as well as in the humans involved in system development, auditing and use. Although we have by no means provided an exhaustive list of examples and cases where diversity and biases exist in intelligent algorithmic systems, we have demonstrated that perspective taking can be an approach to AT.

\section{Towards Algorithmic Transparency}

Having explored the problem space of bias in algorithmic systems, we now describe five levels of \textit{algorithmic transparency}. We shall relate the examples from Section 3 to these levels, to better understand the state of AT research, and to propose directions for future research.

\subsection{First-level AT: User awareness}
In the first level of transparency, which is the most informal, the user becomes aware that something is ``not quite right." The user interprets the system from her own context, $C_i$. However, the context of the developer, $C_D$ is unknown and no formal assessment takes place. 


This form of AT is related to users' digital skills and is documented in the literature. Through repeated interactions with the system, the user can develop a mental model of the system and may learn to be critical of its statements, developing also an awareness of its biases. RS2 described such a case, where users of Booking.com posted reviews, in which they warned others about skewed ratings. Some users may generalize their observations, developing ``folk theories" of how the algorithm works \cite{rader2015understanding}, essentially taking the potential perspective of the developer, building an unformalized, unverified model of $C_D$. However, others show little ability to scrutinize algorithmic statements, particularly those that reinforce their own deeply held beliefs (e.g., concerning gender roles \cite{otterbacher2018investigating}). 

First-level AT represents a very active area of research. However, further work is needed on how to design and deliver user awareness (e.g., explainability  \cite{abdollahi2018transparency}). In other words, future work must address questions such as: what methods are the most efficient for making the user aware of how the system is making its decisions (what data the system holds, how the system uses the data), and/or why specific information is delivered to the user. For example, in \cite{Chen:2018:TAL:3173574.3174111}, the authors conduct a number of experiments to understand how different strategies towards transparency and personalization affect the users. 

\subsection{Second-level AT: Observer audit}
In the second level of transparency, we have an observer who interprets the system's behaviors in terms of $C_V$. As in the first level, the context of the developer/system, $C_D$, is undefined. The observer could be a user of the system; however, in second-level transparency, the observer seeks to evaluate the system for biases, with respect to a cultural consensus on $O_V$, and not that of one particular user ($O_i$). Those in the observer role might be researchers as in the cases in Section 3 and in Andreou et al. \cite{andreou:hal-01955309}, where researchers aimed to understand the effectiveness of Facebook targeted advertising; or even journalists, as in the often-discussed article on racial bias in the COMPAS system for predicting recidivism.\footnote{https://www.propublica.org/article/machine-bias-risk-assessments-in-criminal-sentencing}

As seen in the above mentioned cases, in second-level transparency, the observer is tasked with measuring AB (or more commonly, average AB). She must also define $O_V$ which, as mentioned, can be done by drawing on large corpora from the Web, official published statistics, or even through crowdwork. Since $C_D$ is unknown, the observer conducts an evaluation of the system's empirical behaviors, comparing them to $O_V$, and then documents her observations, as well as the whole process. Thus, the observer's analysis can be replicated by others, and over various points in time, in order to track system behavior.

\subsection{Third-level AT: Developer disclosure}
A third level of transparency is to articulate, at least in part, the context of the developer, $C_D$. We have seen that in first-level transparency, the user may reach a point of ``theorizing" or trying to infer $C_D$, while in second-level transparency, an observer uses empirical experimentation to ``go around" $C_D$. In contrast to these two levels, in third-level transparency, $C_D$ is formalized in the developer's own terms. 
We might relate this level of transparency to the Value Sensitive Design (VSD) approach \cite{friedman2008value}. VSD asks technology developers to consider all stakeholders' values, and how these might be supported or hindered in the system in question (perspective-taking). During this process, a developer would need to first define his own reference context. In a similar vein, professional codes of conduct for developers, ask them to reflect upon their position and the impact of their work. For instance, the ACM Code of Ethics and Professional Conduct\footnote{https://ethics.acm.org/} asks developers to ``be honest and trustworthy" (1.3) and ``be fair and take action not to discriminate" (1.4). Genuine compliance would require self-reflection on / definition of $C_D$ and perspective taking so developer $D$ will minimize biases. 

To articulate $S_D$, a developer would need to describe how knowledge about the world is represented within the system, beginning with the implicit user model on which the system was developed (i.e., who are the intended users). She would also need to document any sources of training data (e.g., how were datasets chosen/built/annotated, and by whom) and resources used by the system (e.g., dictionaries).  Finally, explanations surrounding the treatment of data by algorithmic processes would also be needed (e.g., which features are considered important enough not only to represent in the data, but also to exploit in the model).

To communicate $O_D$, the developer would need to explain what she considers to be the norm or standard. This issue is related again to data (e.g., the choice of training data to represent a given entity) but also to the manner in which the system is trained and evaluated by the developer.\footnote{The above lists of required explanations are not exhaustive.}


\subsection{Fourth-level AT: User/Developer mapping}
 The fourth level of AT assumes that both the first and third levels are feasible. This level allows for the comparison between the reference contexts of the developer and user, $C_D$ and $C_i$. This would include formally mapping the dimensions that each use to represent knowledge of the world (i.e., $S_D$ and $S_i$) as well as mapping their unbiased references (i.e., $O_D$ and $O_i$). A path towards fourth-level AT is for the developer to contextualize user scripts when developing an intelligent algorithmic system, bearing in mind that concepts like ``fairness" are specific to social contexts \cite{selbst2019fairness}. Through this process, the user would be able to understand the perspective of the developer. Such a disclosure may lead to greater trust in the system. Likewise, the system developer could achieve a better understanding of the diversity of the user base, and for whom perceived biases are a problem. These, of course, are value judgments. 
 
 \subsection{Fifth-level AT: Full transparency}
Finally, at the highest level of transparency, any third party observer would be able to see, in full transparency, the positions of the user(s) and the developer. He could then make his own value judgment concerning the system's behaviors, independent of the evaluation of another user and/or the evaluation of the system against a cultural consensus on $O$.

 \subsection{Summary}
The state-of-the-art AT literature primarily documents cases of first- and second-level transparency. That said, efforts towards reaching third-level transparency can be seen both in professional/ethical codes for developers, as well as research efforts that promote the alignment of human values with the manner in which technologies are developed. 

Fourth- and fifth-level AT would be the most difficult to achieve and thus, remain areas for future research. These levels would require a mapping (i.e., projection) of one Cartesian space onto another. As explained in Section 2, the dimensions used to represent a given entity vary across individuals. Furthermore, the dimensions describe different properties (e.g., some are discrete, some may be unordered, etc.) Therefore, such comparisons would become complicated, but could lead to a more general methodology of AT, in which many useful properties might be evaluated (e.g., under which circumstances bias is symmetrical). 

As a final note, we do not mean to imply that it is possible to construct a final, ``objective" reference point, from which an observer can evaluate the world around him. Arguably, the biggest obstacle that we face in working towards algorithmic transparency, is the fact that each one of us lives with our ``bias blind spots." Even if one is able to work through the levels of transparency, documenting his position and perceptions at each step of the process of using/creating a system, it is difficult for us to be aware of all the biases that influence our perception and behaviors \cite{pronin2002bias}. For this reason, we believe that the solution itself is the process of building an ever-changing reference point (and not a final or definitive reference point). 

\section{Conclusion}
In this work, we aimed to bring the issue of \textit{diversity} into the discussion on algorithmic bias and algorithmic system transparency. To this end, we adopted a perspective taking approach from psychology in defining a theoretical framework towards understanding and discussing bias in algorithmic systems, taking into consideration the diversity aspect. Biases and stereotypes occur due to the divergent perspectives, mainly, of the humans involved in the process of designing, developing and using an algorithmic system. Acknowledging diversity is a first step towards appreciating that transparency at any level can contribute to the efforts in reducing biases in algorithmic systems. The developer needs to be in a position to identify who the potential users will be and take their perspectives (e.g., through script writing) when interpreting the system's output. In addition, the perspective of the developer should be made available to the users, when possible, so the user can make informed decisions when receiving an output. 

We have seen that transparency is the obvious solution to the problem of algorithmic system bias. This is because of the diversity of people, and the fact that each one of us has a unique reference context (i.e., perspective), from which we interpret system behaviors and/or from which we develop systems. It is not that we all necessarily see things differently, and thus, disagree as to whether or not a given system is biased. As observed through the examples presented in Section 3, a cultural consensus can typically be established within a broader context (e.g., amongst users who speak the same language, were raised in a similar culture, and/or who have common life experiences). However, the rapid development of Internet technologies in the last decade has led us to a situation in which effectively managing diversity is an everyday phenomenon. 

It can be noticed that, despite that transparency is the definitive solution for bias elimination, full transparency has yet to be achieved. It is complicated by the fact that each one of us is positioned within our own reference context.
The solution to the unreachability of full AT is to make the level of transparency dependent on the nature of the problem and the users' level of digital literacy. The more critical a system, the higher the level of transparency required. In this way, we also minimize the costs, which clearly increase with the level of AT. For instance, for less critical bias problems (e.g., adults using a search engine to find information to aid a decision), it may be sufficient for users to simply be aware of the biases such that they can adjust their actions accordingly. The level of awareness can depend on their digital skills and understanding. In contrast, in cases where the system's decisions might result in more immediate harm (e.g., facial recognition systems processing photos of suspects in a criminal case), it is necessary to insist on the higher level of developer disclosure. 

\subsection{Addendum: on detecting ``fake news"}
In closing, we note the relationship between the perspective on AT advocated in this work, and another pertinent area of research, that of developing algorithms to detect ``fake news." The link between these domains highlights the importance of taking a \textit{diversity perspective} when analyzing problems related to bias, whether the bias manifests in the behaviors of algorithmic systems (AT) or in information itself (fake news). 

Fake news has been characterized as a ``distortion bias on information manipulated by the publisher" \cite{shu2017fake}. \textit{Credibility} is the diversity dimension of interest; research on detecting rumors on social media \cite{qazvinian2011rumor} or classifying posts as to their newsworthiness \cite{castillo2011information} assume that the level of credibility of messages varies. The challenge is to use features of a message, to determine whether or not it deviates significantly from $O$ (a baseline representing credible information). Just as in AT research, incoming observations are evaluated with respect to the observer's reference context ($C_V$). 


In future work, a means to achieve fourth-level AT could also benefit fake news detection. The issue of ``fake news" often enters political debates (e.g., Trump supporters vs. detractors), with each party accusing the other of not acting in good faith (i.e., circulating fake news). However, if the reference contexts of the parties were to be mapped, each could better understand the underlying value judgments of the other. This is parallel to the developer in Section 2, acting in good faith (i.e., the assumption that $PAB_D(a)=0$). Thus, the parties would be positioned to identify genuine ``fake news," differentiating it from that which is simply generated via a reference context that differs from theirs. Of course, to realize this, a diversity perspective is essential.

\section*{Acknowledgements}
This research has been supported by the 
European Union's Horizon 2020 Research and Innovation Program under Grant Agreement No. 810105 (CyCAT: Cyprus Center for Algorithmic Transparency). 

\bibliographystyle{named}
\bibliography{ijcai19}

\end{document}